\documentclass[11pt]{article}
\usepackage[T1]{fontenc}
\usepackage[utf8]{inputenc}
\usepackage{amsmath,amssymb,amsthm}
\usepackage[margin=1in]{geometry}
\usepackage[hidelinks]{hyperref}
\usepackage[title]{appendix}
\numberwithin{equation}{section}
\newtheorem{theorem}{Theorem}[section]
\newtheorem{lemma}[theorem]{Lemma}
\newtheorem{proposition}[theorem]{Proposition}
\newtheorem{corollary}[theorem]{Corollary}
\theoremstyle{remark}
\newtheorem{remark}[theorem]{Remark}

\title{Existence of a third self-similar  solution\\ of the Yang--Mills equation
in high odd dimensions}

\author{%
Piotr Bizo\'n\thanks{Email: \texttt{piotr.bizon@uj.edu.pl}}\,
and Robert Kluszczy\'nski\thanks{Email: \texttt{robert.kluszczynski@student.uj.edu.pl}}\\[1ex]
{\small Institute of Theoretical Physics,
Jagiellonian University, Krak\'ow, Poland}
}

\date{}

\begin{document}
\maketitle
\enlargethispage{2\baselineskip}
\begin{abstract}
In a recent paper~\cite{BGW}, Bizo\'n, Glogi\'c,
and Wasserman (BGW) reduced the classification
of smooth self-similar Yang--Mills profiles in
odd spatial dimensions $d\ge11$ to an algebraic
problem of counting roots of an
explicit polynomial in a prescribed interval. Two 
roots correspond to explicit profiles. BGW conjectured that
there is exactly one additional root and verified this  in a range of odd
dimensions. Here, we prove the existence of an
additional root, and hence of a third smooth
self-similar profile, in every odd dimension
$d\geq11$. The proof combines an endpoint sign calculation for the resonance
polynomial with an algebraic argument showing that the third root is distinct from the two
explicit ones. Uniqueness remains open in general.
\end{abstract}

\section{Introduction}
Continuing the study initiated
in~\cite{BGW}, we consider self-similar solutions of the Yang--Mills equation
in high odd spatial dimensions.
Under the spherically symmetric magnetic ansatz with gauge group $\mathrm{SO}(d)$ the field  equation in $(d+1)$-dimensional Minkowski spacetime reduces to
\begin{equation}\label{eq:phi}
\phi_{tt}=\phi_{rr}+\frac{d-3}{r}\phi_r+\frac{d-2}{r^2}\phi(1-\phi^2).
\end{equation}
The equation is invariant under scaling $\phi(t,r)\mapsto \phi(t/\lambda,r/\lambda)$ and its  conserved energy scales as $\lambda^{d-4}$, hence it is energy-supercritical for $d\geq5$.

A self-similar solution takes the form
\[
\phi(t,r)=u(y),\qquad y=\frac{r}{T-t},
\]
where the profile satisfies the ODE
\begin{equation}\label{profile}
y^2(1-y^2)u''+\bigl((d-3)y-2y^3\bigr)u'+(d-2)u(1-u^2)=0.
\end{equation}
The aim is to construct nonconstant profiles that are smooth
for $0\leq y\leq 1$, corresponding to the causal
past of the blowup point $(T,0)$. Such solutions provide
explicit examples of finite-time singularity formation from smooth 
initial data.  

A first step in the classification of self-similar profiles was taken by Cazenave, Shatah, and Tahvildar-Zadeh \cite{CSSZ}, who showed that for odd $d\geq 5$ smoothness at the
light cone $y=1$ imposes the compatibility condition
\[
c(1-c^2)P_m(c^2)=0,
\]
where $c=u(1)$ and $P_m$ is a recursively constructed polynomial of degree $m=(d-5)/2$.

In their recent paper, Bizo\'n, Glogi\'c, and Wasserman~\cite{BGW}
used a  shooting argument to prove that for $m\geq 3$  each root of
$P_m$ in $(0,1)$  determines a unique (modulo reflection) smooth self-similar profile.
 Their result reduces the classification of these
profiles in odd spatial dimensions $d\geq11$ to the algebraic problem of counting 
roots of $P_m$ in $(0,1)$.

Two distinct roots in $(0,1)$ are given by the squared endpoint
values of the explicit profiles
\begin{equation}\label{intro-explicit}
u_\pm(y)=1-\frac{\alpha_\pm y^2}{y^2+\beta_\pm},
\end{equation}
where
\[
\alpha_\pm=2\pm2\sqrt{\frac{d-4}{3(d-2)}},\qquad
\beta_\pm=\frac23(d-4)\pm\frac13\sqrt{3(d-2)(d-4)}.
\]
BGW conjectured that there is exactly one additional  root, and hence a third  smooth self-similar solution. They verified this conjecture by explicit computation for odd dimensions up to $d=35$.

In this paper, we establish the existence part
of the BGW conjecture by proving the following result.

\begin{proposition}\label{thm:third-root}
For every integer $m\geq3$, the polynomial $P_m$
has a root in $(0,1)$ distinct from the two
explicit roots $u_-(1)^2$ and $u_+(1)^2$,
where $d=2m+5$.
\end{proposition}

The classification theorem of~\cite{BGW} then gives:

\begin{corollary}\label{thm:third}
For every odd dimension $d\geq11$, there are, modulo reflection,
at least three nonconstant
smooth self-similar Yang--Mills profiles,
two of which are given by~\eqref{intro-explicit}.
\end{corollary}

The proof of Proposition~\ref{thm:third-root}
has two main steps. In Section~\ref{sec:endpoints},
we show that $P_m$ has opposite signs at $0$ and $1$.
Factoring out the two explicit roots preserves
these signs, so the remaining factor has a root
in $(0,1)$. To show that this root is distinct
from both explicit ones, we first give a short
irreducibility argument in Section~\ref{sec:third}.
This applies whenever $3(d-2)(d-4)$ is not a
perfect square and, in particular, covers
$3\leq m\leq11$. The remaining, exceptional
dimensions form a sparse sequence beginning
with $d=29,365,5045,\ldots$. To handle these cases,
we prove in Section~\ref{sec:simplicity} that
the explicit roots are simple for all $m\geq12$
and hence cannot be roots of the remaining factor.

\section{The resonance polynomial and its endpoint signs}\label{sec:endpoints}
Introducing
\[
x=\frac1y,\qquad v(x)=u(1/x),
\]
we rewrite equation~\eqref{profile} as
\begin{equation}\label{inverse}
(x^2-1)v''-2mxv'+(2m+3) v(1-v^2)=0.
\end{equation}
Inserting the formal expansion,
\begin{equation}\label{taylor}
v(x;c)=\sum_{n\geq0}a_n(c)(x-1)^n,\qquad a_0(c)=c,
\end{equation}
into \eqref{inverse} and setting the coefficient of each  power of  $(x-1)$  to zero gives the recurrence relation
\begin{equation}\label{nonlinearrec}
2(n+1)(n-m)a_{n+1}+[n^2-(2m+1)n+(2m+3)]a_n
-(2m+3)\sum_{\substack{i,j,k\geq0\\i+j+k=n}}a_i a_j a_k=0.
\end{equation}
This recurrence determines   the coefficients $a_1(c),\dots,a_m(c)$ successively from $a_0=c$. At $n=m$ the coefficient of $a_{m+1}$ vanishes, leaving the compatibility condition $K_m(c)=0$, where
\begin{equation}\label{residual}
K_m(c)=[-m(m+1)+(2m+3)]a_m(c)-(2m+3)\sum_{i+j+k=m}a_i(c)a_j(c)a_k(c).
\end{equation}
Thus $K_m(c)=0$ is precisely the condition for continuing
the Taylor expansion through the resonant order.
If this condition holds, then $a_{m+1}$ remains free.

The recurrence shows that $K_m$ is an odd polynomial
with rational coefficients and vanishes at $c=0,\pm1$.
Hence $K_m(c)/[c(1-c^2)]$ is a polynomial in $c^2$.
We define the resonance polynomial $P_m$ by
\begin{equation}\label{normalization}
K_m(c)=(-1)^m c(1-c^2)P_m(c^2).
\end{equation}
The factor $(-1)^m$ is included for later convenience.
The polynomial $P_m$ agrees, up to a nonzero
constant factor, with the resonance polynomial of~\cite{BGW}.

\begin{proposition}\label{endpoints}
For every integer $m\geq3$, the polynomial $P_m$  satisfies
\[
P_m(0)<0<P_m(1).
\]
\end{proposition}
\begin{proof}
For $c=0$ or $c=1$ the solution is constant.
Thus, fixing $\eta\in\{0,1\}$, we have
\[
a_0(\eta)=\eta,\qquad
a_n(\eta)=0\quad(1\leq n\leq m).
\]
Set $b_n:=a_n'(\eta)$, so that $b_0=1$.
Differentiating~\eqref{nonlinearrec} and evaluating
at $c=\eta$ gives, for $0\leq n<m$,
\[
2(n+1)(n-m)b_{n+1}
+\bigl[n^2-(2m+1)n+(2m+3)(1-3\eta^2)\bigr]b_n=0,
\]
hence
\[
b_{n+1}
=\frac{n^2-(2m+1)n+(2m+3)(1-3\eta^2)}
       {2(n+1)(m-n)}b_n.
\]
Iterating from $b_0=1$, we obtain
\[
\begin{aligned}
b_m
&=\prod_{n=0}^{m-1}
\frac{n^2-(2m+1)n+(2m+3)(1-3\eta^2)}
     {2(n+1)(m-n)}\\
&=\frac1{2^m(m!)^2}
\prod_{n=0}^{m-1}
\bigl[n^2-(2m+1)n+(2m+3)(1-3\eta^2)\bigr].
\end{aligned}
\]

Differentiating~\eqref{residual} at $c=\eta$ gives
\[
K_m'(\eta)
=\bigl[-m(m+1)+(2m+3)(1-3\eta^2)\bigr]b_m.
\]
The factor in brackets is precisely
$n^2-(2m+1)n+(2m+3)(1-3\eta^2)$ evaluated at $n=m$.
It therefore supplies the final factor in the product:
\begin{equation}\label{Kprimeproduct}
K_m'(\eta)=\frac1{2^m(m!)^2}
\prod_{n=0}^{m}
\bigl[n^2-(2m+1)n+(2m+3)(1-3\eta^2)\bigr].
\end{equation}

\paragraph{Evaluation at $c=0$.}
Differentiating~\eqref{normalization} at $c=0$ gives
\[
P_m(0)=(-1)^m K_m'(0).
\]
Setting $\eta=0$ in~\eqref{Kprimeproduct}, we obtain
\[
P_m(0)=\frac{(-1)^m}{2^m(m!)^2}
\prod_{n=0}^{m}
\bigl[n^2-(2m+1)n+2m+3\bigr].
\]
The factors at $n=0$ and $n=1$ are $2m+3$ and $3$,
respectively. The remaining $m-1$ factors are negative,
since the quadratic is decreasing on $[2,m]$ and
its value at $n=2$ is $5-2m<0$.
The product therefore has sign $(-1)^{m-1}$, and hence
\[
P_m(0)<0.
\]

\paragraph{Evaluation at $c=1$.}
Differentiating~\eqref{normalization} at $c=1$ gives
\[
K_m'(1)=(-1)^{m+1}2P_m(1).
\]
For $\eta=1$, the factors in~\eqref{Kprimeproduct}
simplify to
\[
n^2-(2m+1)n-2(2m+3)=(n+2)(n-2m-3),
\]
giving
\[
\begin{aligned}
K_m'(1)
&=\frac1{2^m(m!)^2}
\prod_{n=0}^{m}(n+2)(n-2m-3)\\
&=(-1)^{m+1}\frac{(2m+3)!}{2^m(m!)^2}.
\end{aligned}
\]
Thus
\[
P_m(1)=\frac{(2m+3)!}{2^{m+1}(m!)^2}>0.
\]
\end{proof}
\section{A third root}\label{sec:third}

The explicit solutions~\eqref{intro-explicit} yield the two roots
$z_\pm=u_\pm(1)^2$, given by
\begin{equation}\label{eq:z}
z_\pm=
\frac{d^3-16d^2+89d-146
\pm4(d-7)\sqrt{3(d-2)(d-4)}}{(d-2)(d-1)^2}.
\end{equation}
Factoring out these roots, we write
\[
P_m(z)=Q_m(z)S_m(z),\qquad Q_m(z)=(z-z_-)(z-z_+).
\]
Since $0<z_\pm<1$, the quadratic factor $Q_m$ is positive at both endpoints.
Proposition~\ref{endpoints} therefore gives
$S_m(0)<0<S_m(1)$, so $S_m$ has a root in $(0,1)$ by the
intermediate value theorem. 

To complete the proof of Proposition~\ref{thm:third-root},
it remains to show that the root of $S_m$ is
distinct from both explicit roots. We divide the proof into two cases.
First, we treat the nonexceptional dimensions,
in which $3(d-2)(d-4)$ is not a perfect square,
using an irreducibility argument.
We then handle the exceptional dimensions
using simplicity of the explicit roots.

\subsection{Nonexceptional dimensions}

Suppose that $3(d-2)(d-4)$ is not a perfect square.
Then $Q_m$ is irreducible over the field $\mathbb{Q}$
of rational numbers.
Write
\[
P_m=Q_m^kR_m,
\]
where $R_m$ has rational coefficients and $k\geq1$
is maximal.
Neither $z_-$ nor $z_+$ is a root of $R_m$:
otherwise the irreducibility of $Q_m$ over $\mathbb{Q}$
would imply that $Q_m$ divides $R_m$, contradicting
the maximality of $k$.
Since $Q_m(0)>0$ and $Q_m(1)>0$, Proposition~\ref{endpoints} implies
\[
R_m(0)<0<R_m(1).
\]
Thus $R_m$ has a root in $(0,1)$ distinct from both
explicit roots, proving Proposition~\ref{thm:third-root}
in every nonexceptional dimension.
In particular, this covers $3\leq m\leq11$.

\subsection{Exceptional dimensions}

The exceptional dimensions are those for which
$3(d-2)(d-4)$ is a perfect square, which is equivalent to the Pell equation
\[
(d-3)^2-3k^2=1
\]
for some integer $k$. 
Among odd dimensions $d\geq11$, this yields the sequence 
$d=29,365,5045,70229,\ldots$. Thus every exceptional dimension
under consideration satisfies $m\geq12$.
To handle these cases, we use the following general
result. Since its proof may be of independent interest,
we present it separately in Section~\ref{sec:simplicity}.

\begin{proposition}\label{prop:simple}
For every $m\geq12$, the two explicit roots
of $P_m$ are simple, i.e. $P_m'(z_\pm)\neq0$.
\end{proposition}

It follows that $S_m(z_\pm)\neq0$ for $m\geq 12$. Thus the root of $S_m$ in $(0,1)$ is distinct
from both explicit roots, completing the proof
of Proposition~\ref{thm:third-root}.

\section{Simplicity of the explicit roots}\label{sec:simplicity}
We prove simplicity of the two explicit roots for $m\geq12$
by showing that the corresponding linearized resonance obstructions
are nonzero. The argument uses coefficient inequalities established
inductively from a three-term recurrence. The initial inequalities
are verified in Appendix~\ref{app:initialization}.
Throughout, we assume that $m\geq12$, set $d=2m+5$, and write
$r=\sqrt{3(d-2)(d-4)}$.

\subsection{Linearized resonance and simplicity}

For either solution branch~\eqref{intro-explicit}, the explicit profile in the coordinate $x=1/y$ is
\[
v(x)=1-\frac{\alpha}{1+\beta x^2},
\]
where
\[
\alpha=2\pm\frac{2r}{3(d-2)},\qquad
\beta=\frac{2(d-4)\pm r}{3}.
\]
We write $c_\pm=v_\pm(1)$, so that $c_\pm^2=z_\pm$ and  suppress the branch subscripts  on $v$, $\alpha$, and $\beta$ 
to avoid cluttering the notation.

Linearizing~\eqref{inverse} about $v$ we obtain
\begin{equation}\label{app:linearized}
(x^2-1)w''-2mxw'+(d-2)(1-3v^2)w=0.
\end{equation}
Let
\begin{equation}\label{app:change}
s=\frac{1-x^2}{1+\beta x^2},
\qquad
w(x)=(1+\beta s)g(s).
\end{equation}
Since
\[
s(1)=0,\qquad s'(1)=-\frac{2}{1+\beta}\neq0,
\]
this is an invertible analytic change of variable near the light cone.
Using
\[
(d-2)\alpha^2=8(1+\beta),
\qquad
q=\frac{3(d-2)\pm r}{2}
  =\frac{3\beta+2m+7}{2},
\]
equation~\eqref{app:linearized} becomes
\begin{equation}\label{app:geq}
g''+
\left(
-\frac{m}{s}
+\frac{\beta(m+\frac72)}{1+\beta s}
+\frac{1}{2(s-1)}
\right)g'
+\frac{q-4-4\beta s}{s(s-1)(1+\beta s)}\,g=0.
\end{equation}
Its indicial roots at $s=0$ are $0$ and $m+1$.
For the exponent-zero expansion, write
\[
g(s)=\sum_{n\geq0}b_ns^n,\qquad b_0=1,\qquad b_{-1}=0.
\]
Substitution into~\eqref{app:geq} gives
\begin{equation}\label{eq:brec}
(n+1)(m-n)b_{n+1}+A_nb_n+C_nb_{n-1}=0,
\qquad 0\leq n<m,
\end{equation}
where
\[
A_n=(n-1)\left(n-m+\frac12\right)
-\frac{\beta}{2}(2n-1)(n+3),
\qquad
C_n=\beta(n+3)(n-2).
\]
The recurrence uniquely determines $b_1,\ldots,b_m$.
At $n=m$, the coefficient of $b_{m+1}$ vanishes, leaving the
compatibility condition $\mathcal E_m=0$, where
\begin{equation}\label{eq:obstruction}
\mathcal E_m=A_mb_m+C_mb_{m-1}.
\end{equation}
The coefficients $b_n$ and the obstruction $\mathcal E_m$ are
understood to depend on the chosen branch.

\begin{lemma}\label{lem:equivalence}
For either  branch, the explicit root is simple if and only if
the corresponding obstruction $\mathcal E_m$ is nonzero.
\end{lemma}

\begin{proof}
Differentiating the nonlinear endpoint recurrence at $c=c_\pm$
and applying~\eqref{app:change}, we obtain
\begin{equation}\label{eq:obstruction-derivative}
\mathcal E_m
=\left(\frac{1+\beta}{2}\right)^{m+1}
z_\pm(1-z_\pm)P_m'(z_\pm).
\end{equation}
The prefactor is positive, proving the equivalence.
\end{proof}

\subsection{Propagation of the coefficient inequalities by induction}

We rewrite~\eqref{eq:brec} as
\begin{equation}\label{eq:difference}
D_n(b_{n+1}-b_n)
=G_nb_n+C_n(b_n-b_{n-1}),
\end{equation}
where
\begin{equation}\label{eq:gap}
D_n=(n+1)(m-n),\qquad
G_n=-A_n-D_n-C_n
=\frac{3(\beta+1)n+9\beta-4m+1}{2}.
\end{equation}
The initial inequalities, verified in
Appendix~\ref{app:initialization}, are $G_8>0$ and
\begin{equation}\label{eq:initial}
\begin{cases}
b_8<b_7<0, & \text{for the plus branch},\\
b_8>b_7>0, & \text{for the minus branch}.
\end{cases}
\end{equation}
For either branch,
\[
G_{n+1}-G_n=\frac32(\beta+1)>0.
\]
Together with $G_8>0$, this implies $G_n>0$ for
$8\leq n\leq m$. Hence equation~\eqref{eq:difference} gives,
for $8\leq n<m$,
\[
\begin{aligned}
b_n<b_{n-1}<0
&\quad\Longrightarrow\quad b_{n+1}<b_n<0,\\
b_n>b_{n-1}>0
&\quad\Longrightarrow\quad b_{n+1}>b_n>0.
\end{aligned}
\]
Starting from~\eqref{eq:initial}, induction yields
\[
\begin{cases}
b_m<b_{m-1}<0, & \text{for the plus branch},\\
b_m>b_{m-1}>0, & \text{for the minus branch}.
\end{cases}
\]
Since $D_m=0$, the definition of $G_m$ gives
\[
\mathcal E_m
=A_mb_m+C_mb_{m-1}
=-G_mb_m-C_m(b_m-b_{m-1}).
\]
Thus $\mathcal E_m>0$ for the plus branch and
$\mathcal E_m<0$ for the minus branch.
By~\eqref{eq:obstruction-derivative}, for $m\geq12$
\[
P_m'(z_+)>0,\qquad P_m'(z_-)<0,
\]
which proves  Proposition~\ref{prop:simple}.
\begin{remark}
Simplicity of $z_+$ also follows from the absence of a zero
eigenvalue for the linearized perturbations about $u_+$, established
by Glogi\'c~\cite{Glogic}. Indeed, if the resonance obstruction
vanished, both local solutions at the light cone would be analytic,
so the zero-mode solution regular at the origin would also be
smooth at the light cone. We have given a direct proof to keep
the exposition self-contained and to treat both branches uniformly.
\end{remark}

\section{Open problems}\label{sec:open-problems}

Having established the existence part of the BGW conjecture,
we conclude with three open problems.

\paragraph{Uniqueness.}
The remaining part of the BGW conjecture is to prove that $P_m$
has exactly three distinct roots in $(0,1)$ for every $m\geq3$.
Using AI-assisted exact root-counting computations, we have
verified this for odd dimensions $11\leq d\leq101$, but a proof
of uniqueness for all odd $d\geq11$ remains open.

\paragraph{Even dimensions.}
Although our proof applies only to odd dimensions, numerical
computations suggest that the third profile persists as $d$
varies continuously, just as the two explicit profiles do.
We therefore conjecture that a third smooth self-similar profile
exists in every  dimension $d\geq11$.

In even dimensions, a local solution near the light cone generally
contains a term with a noninteger power, whose coefficient
must vanish for the solution to be smooth.
Unlike the logarithmic coefficient in odd dimensions, this
coefficient is not determined by the endpoint value but is
an independent local parameter.
For a solution regular at the origin, its value must be
determined by continuation to the light cone.
Although existence is a connection problem in both odd and
even dimensions, the shooting theorem of~\cite{BGW} reduces
the odd-dimensional case to an algebraic condition on the
profile value $u(1)$ at the past light cone.
No analogous reduction is available in even dimensions,
where one must instead control the coefficient of the
noninteger-power term along the family of solutions
regular at the origin.
We shall address this connection problem in a forthcoming paper.

\paragraph{Stability and dynamics.}
The rigorous nonlinear stability results~\cite{Do,Glogic},
combined with the numerical evidence~\cite{BT,BB}, indicate
that the self-similar profile $u_+$ governs generic spherically
symmetric blowup.
The dynamical roles of $u_-$ and $u_\ast$, on the other hand,
remain to be understood.
For odd dimensions, the numerical spectral analysis
reported in~\cite{BGW} indicates that, apart from the
time-translation mode, $u_-$ and $u_\ast$ have two and one
unstable modes, respectively, for $11\leq d\leq17$, with
these counts reversed for $d\geq19$.
The profile with a single unstable mode is expected to act as a
critical solution whose codimension-one stable manifold separates
blowup from dispersion.

A first step toward understanding these dynamics is to establish
the spectral properties of $u_-$ and $u_\ast$ rigorously.
For the explicit profile $u_-$, the spectral problem reduces
to a connection problem for the Heun equation, analogous to
that studied for $u_+$ in~\cite{BB}.
Treating $d$ as a real parameter, numerical computations indicate
that $u_-$ and $u_\ast$ meet at $d\approx17.22278$ in what
appears to be a transcritical bifurcation, exchanging one
unstable mode.
The Heun formulation provides a route to locating the associated
zero eigenvalue on the explicit branch and establishing its
crossing through zero.
For $u_\ast$  a different approach
is required, since the profile is not known in closed form.

\begin{appendices}
\section{Verification of the initial inequalities}
\label{app:initialization}
For $m\geq 12$, we establish the initial inequalities used in
Section~\ref{sec:simplicity}:
\[
G_8>0,
\qquad
\begin{cases}
b_8<b_7<0, & \text{for the plus branch},\\
b_8>b_7>0, & \text{for the minus branch}.
\end{cases}
\]
Exact arithmetic calculations in this appendix were performed using Maple.
\subsection{Parametrization and coefficient formulas}
Set
\[
h=
\begin{cases}
-\dfrac{r}{d-2}, & \text{for the plus branch},\\[4pt]
 \dfrac{r}{d-2}, & \text{for the minus branch}.
\end{cases}
\]
Then
\begin{equation}\label{app:param}
m=\frac{3(h^2-1)}{2(3-h^2)},
\qquad
\beta=\frac{2h(2h-3)}{3(3-h^2)}.
\end{equation}
Since
\[
h^2=\frac{3(2m+1)}{2m+3},
\]
the condition $m\geq12$ gives the ranges
\[
-\sqrt3<h\leq-\frac53
\quad\text{for the plus branch},
\qquad
\frac53\leq h<\sqrt3
\quad\text{for the minus branch}.
\]
Substitution into~\eqref{eq:gap} yields on both intervals
\[
2G_8=33\beta-4m+25
=-\frac{(h-3)(13h-27)}{h^2-3}>0.
\]

Starting from $b_0=1$ and $b_{-1}=0$, successive application
of~\eqref{eq:brec} gives
\begin{equation}\label{app:b7formula}
b_7=
-\frac{1024h^2(h-3)(2h-3)^2U(h)}{D(h)}
\end{equation}
and
\begin{equation}\label{app:diff78}
b_8-b_7=
-\frac{2048h^2(h-3)^2(2h-3)^2V(h)}
{3(17h^2-45)D(h)},
\end{equation}
where
\[
\begin{aligned}
D(h)&=25515(h^2-1)(3h^2-7)(5h^2-13)(5h^2-9)(7h^2-15)(11h^2-27)(13h^2-33),\\[4pt]
U(h)&=700700h^9-3114034h^8+3854364h^7-495147h^6+509976h^5\\
&\quad-1651644h^4-2119770h^3+380295h^2+1786050h+765450,\\[4pt]
V(h)&=5955950h^{10}-39919264h^9+97285422h^8-96662700h^7+25308045h^6\\
&\quad-1859841h^5+12088278h^4+11565585h^3-3433590h^2-9185400h-3444525.
\end{aligned}
\]
Every factor in $D(h)$, as well as $17h^2-45$, is strictly
positive for $25/9\leq h^2<3$.

\subsection{The plus branch}

Put
\[
t=-h-\frac53\geq0.
\]
Direct expansion shows that both
\[
-U\!\left(-\frac53-t\right)
\quad\text{and}\quad
V\!\left(-\frac53-t\right)
\]
have strictly positive coefficients as polynomials in $t$.
Hence $U(h)<0$ and $V(h)>0$ on the plus-branch interval.
Since $h-3<0$, equations~\eqref{app:b7formula}
and~\eqref{app:diff78} imply
$
b_7<0$ and
$b_8-b_7<0$, hence
$b_8<b_7<0$.

\subsection{The minus branch.}
Put $t=h-5/3$, so that $0\leq t<\sqrt3-5/3<1/15$.
Direct expansion shows that $U(5/3+t)$ has strictly positive
coefficients, and hence $U(h)>0$ on the minus-branch interval.

To determine the sign of $V(h)$, write
\[
F(t)=V\!\left(\frac53+t\right)=\sum_{j=0}^{10}a_jt^j.
\]
Expansion gives
\[
a_0,a_1,a_2<0,\qquad a_j>0\quad(3\leq j\leq10).
\]
Dropping the nonpositive linear and quadratic terms and bounding
the higher-order terms at $t=1/15$, we obtain, for $0\leq t\leq1/15$,
\[
F(t)
\leq a_0+\sum_{j=3}^{10}\frac{a_j}{15^j}
=F\!\left(\frac1{15}\right)
  -\frac{F'(0)}{15}
  -\frac{F''(0)}{450}
=-\frac{4074530128229}{38443359375}<0.
\]
Thus $V(h)<0$ throughout the minus-branch interval.
Equations~\eqref{app:b7formula} and~\eqref{app:diff78}, together
with $h-3<0$, yield $b_7>0$ and $b_8-b_7>0$,
$b_8>b_7>0$.

\end{appendices}

\end{document}